# A Universal, Rapid Method for Clean Transfer of Nanostructures onto Various Substrates[**]

*Hai Li,[1] Jumiati Wu,[1] Xiao Huang,[1] Zongyou Yin,[1] Juqing Liu,[1] Hua Zhang[1,*]*

[1]School of Materials Science and Engineering, Nanyang Technological University, 50 Nanyang Avenue, Singapore 639798, Singapore.

*To whom correspondence should be addressed.

Phone: +65-6790-5175. Fax: +65-6790-9081

E-mail: hzhang@ntu.edu.sg

Website: http://www.ntu.edu.sg/home/hzhang/

**ABSTRACT.** Transfer and integration of nanostructures onto target substrates is the prerequisite for their fundamental studies and practical applications. Conventional transfer techniques that involve stamping, lift-off and/or striping suffer from the process-specific drawbacks, such as the requirement for chemical etchant or high-temperature annealing and the introduction of surface discontinuities and/or contaminations that can greatly hinder the properties and functions of the transferred materials. Herein, we report a universal and rapid transfer method implementable at mild conditions. Nanostructures with various dimensionalities (*i.e.* nanoparticles, nanowires and nanosheets) and surface properties (*i.e.* hydrophilic and hydrophobic) can be facilely transferred to diverse substrates including hydrophilic, hydrophobic and flexible surfaces with good fidelity. Importantly, our method ensures the rapid and clean transfer of two-dimensional (2D) materials, and allows for the facile fabrication of vertical heterostructures with various compositions used for electronic devices. We believe that our method can facilitate the development of nano-electronics by accelerating the clean transfer and integration of low-dimensional materials into multidimensional structures.

Transfer of nanostructures from one substrate to another is important for large-area integration of nanoscale building blocks into functional structures with multidimensionality for a wide scope of applications.[1] Examples of previously reported transfer techniques include kinetic-controlled transfer printing with poly(dimethyl siloxane) (PDMS) stamp,[2] lift-off by etching,[3] striping/peeling mediated by poly(methyl methacrylate) (PMMA),[4-6] and water-assisted wedging[7] or peel-and-stick method.[8] However, these methods are not generally applicable as they suffer from the process-specific drawbacks, such as the requirement for the dedicate control over the adhesion force at different interfaces,[2] intolerance of transferred nanostructures to chemical etchant[3, 8] and the harsh thermal environment needed for complete removal of polymer residues,[4, 9-10] as well as the wet-process induced wrinkling of sheet structures.[7, 11-12] Importantly, the defects resulting from the transfer process become especially detrimental to the performance of devices made from delicate structures at nanoscale. For example, the transfer of two-dimensional (2D) nanomaterials, such as graphene and transition metal dichalcogenide (TMD) nanosheets, has been realized by various methods such as those based on the PMMA mediator,[9-10, 12-17] thermal-release tape[18] and "self-release layer".[19] However, the structural discontinuities in the nanosheets, such as cracks and wrinkles,[10, 12, 14-15] and the polymer residues that could not be completely removed by washing at mild conditions[9, 14] have dramatically hindered their performance in electronic and optoelectronic devices.[11] Therefore, development of a general method for clean and precise-fidelity transfer of nanostructures is highly desirable.

Herein, we report a universal, rapid method to transfer various low-dimensional nanomaterials, including zero-dimensional (0D) nanoparticles (NPs), one-dimensional (1D) nanowires (NWs), and 2D nanosheets (graphene and TMD nanosheets) as well as their hybrid structures from the $SiO_2$/Si substrate onto diverse substrates with good fidelity. The whole transfer process, including peeling off, transfer and cleaning, carried out at mild conditions can be finished within several minutes. No structural defect is observed on the transferred low-dimensional nanomaterials.

Importantly, by using our method, the heterostructures of graphene-MoS$_2$, graphene-WSe$_2$, MoS$_2$-WSe$_2$ and MoS$_2$-TaS$_2$ have also been easily prepared.

**RESULTS AND DISCUSSION**

Figure 1 schematically shows the transfer process of various low-dimensional nanomaterials (see Materials and Methods for details). Briefly, a polymer film (*e.g.* poly(L-lactic acid), PLLA) as carrier is spin-coated on the low-dimensional nanomaterials (*e.g.* 0D NPs, 1D NWs or 2D nanosheets) pre-deposited on a freshly cleaned Si substrate coated with 90 nm SiO$_2$, referred to as 90 nm SiO$_2$/Si substrate (Figure 1a,b). About 1-mm-wide polymer strips at the edges of SiO$_2$/Si substrate are then scratched off to expose the hydrophilic SiO$_2$ surface (Figure 1c). After a thick PDMS film is brought into conformal contact with the polymer film (Figure 1d), a small water droplet (~30 μL), which is introduced at one exposed hydrophilic SiO$_2$/Si edge, penetrates between the hydrophilic SiO$_2$/Si substrate and hydrophobic carrier polymer attached with nanomaterials, and immediately separates them within several seconds (Figure 1e,f, and Video S1 in Supporting Information (SI)). After the PDMS/polymer/nanomaterial film is peeled off and then brought into conformal contact with the target substrate, the PDMS film is peeled off on a hotplate at 50 °C (Figure 1g,h, and Video S2 in SI). The carrier polymer is then completely removed by being dissolved in dichloromethane (DCM) at 50 °C within several tens of seconds (Video S3 in SI), and the nanomaterials are left on the target substrate (Figure 1i). Here, it is noteworthy that in the previously reported transfer of nanomaterials which were grown on SiO$_2$/Si, the etching of SiO$_2$ to peel off PMMA with attached nanomaterials usually took more than 30 min in the hot KOH or NaOH solution, which is time-consuming and not environment-friendly.[16-17] In addition, annealing at high temperature was required to completely remove the carrier polymer (*i.e.* PMMA).[10, 17] In contrast, the whole process of our transfer method, including polymer deposition, peeling off, transfer and cleaning, is carried out at mild conditions and can be finished within several minutes.

By using our method, the transfer of 0D and 1D nanomaterials are shown in Figures 2 and 3. The dot array of Au NPs adsorbed on an aminosilane-patterned 90 nm $SiO_2$/Si substrate[20] was transferred to a 300 nm $SiO_2$/Si substrate (Figure 2a,b). By comparing the optical microscopy (OM) images of the Au NPs array on the original (Figure 2a) and target substrate (Figure 2b), we can evidently observe the high fidelity of our transfer process as indicated by the red and blue arrows which point to the same position in the Au NPs array before and after the transfer, respectively. Importantly, from the zoom-in atomic force microscope (AFM) images (insets in Figure 2a,b, and Figure 3), it can be clearly seen that all individual Au NPs maintained their own positions after the transfer. In addition, 1D nanomaterials, *i.e.* silver nanowires (Ag NWs) used here, were also completely transferred from 90 nm $SiO_2$/Si onto 300 nm $SiO_2$/Si with good fidelity (Figure 2c,d).

Currently, one of the most important applications that demand clean and facile transfer techniques is the 2D nanosheet-based device.[21-26] Impressively, our method can be used for the wrinkle-free transfer of various 2D nanomaterials on various substrates. As shown in Figure 4, single-layer (1L) graphene (Figure 4a,b), quadruple-layer (4L) $MoS_2$ (Figure 4g,h) and double-layer (2L) $WSe_2$ (Figure 4m,n) nanosheets were successfully transferred from 90 nm $SiO_2$/Si onto 300 nm $SiO_2$/Si. There is no wrinkle or crack observed on these 2D nanosheets from both the OM and AFM characterizations (Figure 4c and d, i and j, and o and p). Importantly, there is no height change for the graphene, $MoS_2$ and $WSe_2$ nanosheets after they are transferred (Figure 4e and f, k and l, and q and r), indicating that DCM at 50 °C can considerably remove PLLA residue on the nanosheets and target substrate. The root-mean-square (rms) surface roughness ($R_a$) of the 2D nanosheets is also similar before and after the transfer (Table 1). For example, the $R_a$ values of 1L graphene nanosheet before and after the transfer are 0.21 and 0.17 nm, respectively, which are much smaller than that of graphene transferred by the common PMMA method and cleaned with acetone ($R_a$ = 0.54 nm,[27] see Table 1 and Figure 5), and comparable to that of graphene treated with thermal annealing at 300 °C ($R_a$ = 0.20 nm).[27] Raman spectroscopy was also used to characterize these 2D nanosheets (Figure S1 in SI), and showed no peak shift before and after the transfer. The transfer of 2D nanosheets with good fidelity was further confirmed by fabrication of field-effect transistors (FETs) based on the pristine and transferred $MoS_2$ nanosheets, which showed comparable performance (Figure S2 in SI).

In our transfer method, besides PLLA, other polymers, such as PMMA and L-lactide-ε-caprolactone copolymer (PLC), can also be used as the carrier, which can be significantly removed from the transferred nanostructures after being cleaned in DCM at 50 °C (Figures S3-4 in SI). In addition, our method is applicable for transfer of nanosheets with different surface properties such as hydrophobic nanosheets (*e.g.* graphene, $MoS_2$ and $WSe_2$, Figure 4) and hydrophilic nanosheets (*e.g.* graphene oxide (GO) and mica, Figure S5 in SI). Moreover, the target substrate is not limited to $SiO_2$/Si, and we have also successfully demonstrated the clean transfer of $MoS_2$ nanosheets to diverse substrates, such as flexible polymer (poly(ethylene terephthalate) (PET)) film (Figure 6), hydrophobic substrate (octadecyltrichlorosilane (OTS) modified $SiO_2$/Si) (Figure S6 in SI), single crystal substrates ($BiFeO_3$, $LiNbO_3$ and PMN-PT) (Figure S7 in SI), indicating the versatility of our method for transferring low-dimensional nanomaterials for various potential applications.

More significantly, our method can also be used for the facile and high-fidelity transfer of complex hybrid structures. For example, graphene nanosheet was firstly deposited on 90 nm $SiO_2$/Si by using the mechanical exfoliation method (Figure S8a in SI). Gold nanoparticles (Au NPs) adsorbed on aminosilane-patterned 90 nm $SiO_2$/Si substrate[20] were transferred onto the graphene nanosheet to form the Au NP-graphene composite by using our clean transfer method (Figure 7a and Figure S8b in SI). As shown in Figure 7a-d and Figure S8 in SI, the composite of Au NPs and graphene nanosheet, referred to as Au NP-graphene, was simultaneously transferred from the original substrate to target substrate without losing any surface feature. Magnified scanning electron microscope (SEM) images show that each Au NP was transferred with good fidelity (Figure 7e and f). In addition, the Ag NW-GO can also be completely transferred with high fidelity at single NW resolution (Figure 7g and h).

The key advantages of our transfer method over previous techniques, *i.e.* rapidness, cleanness and high precision, are especially important in fabrication of vertically stacked heterostructures of graphene and other 2D nanosheets, which have exhibited promising optical and electronic properties that ravel their individual components.[28-37] Compared to the common $SiO_2$-supported graphene or $MoS_2$ field-effect transistors (FETs), devices produced from graphene or $MoS_2$ stacked on hexagonal boron nitride (h-BN) have shown enhanced mobility and ON-OFF ratio by roughly an order of magnitude.[28-29] Other high-performance devices, such as field-effect tunneling transistors, logic transistors and memory devices, have also been fabricated based on the heterostructures of graphene-$MoS_2$/$WS_2$.[29-30, 32, 34-36] Previously, to make such vertical heterostructures, different 2D nanosheets were transferred layer by layer in sequence by using the common "dry" transfer method based on PMMA.[28-29] In order to remove the polymer residue, after each transfer step, the polymer must be dissolved in acetone and annealed (*e.g.* in a mixture of $H_2$/Ar at 250 ºC) before the subsequent transfer of another layer. However, the PMMA residue cannot be completely removed even with high-temperature annealing.[9] Impressively, our method is able to considerably remove the polymer residue in DCM at 50 ºC without any annealing process, which greatly shortens the fabrication time required for the multi-step transfer. As an example, FET devices were fabricated based on the heterostructures of graphene and $MoS_2$ by using our method (graphene-$MoS_2$ FET) (Figure S9 in SI). The heterostructure of graphene and $MoS_2$ was prepared by transferring a triple-layer (3L) $MoS_2$ nanosheet onto a 3L graphene nanosheet (Figure S9 in SI). Figure 8a shows the $I_d$-$V_d$ curves of the FET device based on $MoS_2$ nanosheet ($MoS_2$ FET) with the Au/Cr electrodes located on the $MoS_2$ nanosheet (electrodes 1 and 2 shown in Figure 8d). Figure 8b shows the $I_d$-$V_d$ curves of the FET device based on the heterostructure of graphene and MoS$_2$ (graphene-MoS$_2$ FET) with the Au/Cr electrodes located on the graphene (electrode 4 shown in Figure 8d) and MoS$_2$ (electrode 2 shown in Figure 8d) nanosheets, respectively. The graphene-MoS$_2$ FET shows much higher I$_d$ compared to MoS$_2$ FET at the backgate voltage (V$_g$) ranging from 0 to 40 V (Figure 8a-b). When the drain voltage (V$_d$) is 2V, the I$_d$ of graphene-MoS$_2$ FET (~19 nA) is about three orders of magnitude higher than that of MoS$_2$ FET (~22 pA) at V$_g$ of 40 V. Figure 8c shows the I$_d$-V$_g$ curves of the transferred graphene-MoS$_2$ FET and MoS$_2$ FET. It is obvious that the graphene-MoS$_2$ FET shows much higher ON current than that of MoS$_2$ FET. The ON/OFF ratio of the graphene-MoS$_2$ FET (~6.3×10$^4$) is about 50 times higher than that of the MoS$_2$ FET (~1.2×10$^3$), indicating the heterostructure of graphene-MoS$_2$ shows much better performance than does the MoS$_2$ nanosheet.

Highly photoresponsive transistors based on MoS$_2$ nanosheets have been reported previously.[24,38-39] Here, we investigated the photoresponse of the heterostructure of graphene-MoS$_2$ (graphene-MoS$_2$ FET) and compared it with the pure MoS$_2$ (MoS$_2$ FET). For both the graphene-MoS$_2$ FET and MoS$_2$ FET, as the incident optical power (P$_{light}$) increases from 2.7 μW to 3.9 mW, the photocurrent (I$_{ph}$), *i.e.* I$_d$ shown in Figure 9a and Figure S10 in SI, increases gradually. As shown in Figure 9a, both devices show similar current in the dark environment. However, the graphene-MoS$_2$ FET has I$_{ph}$ about two orders higher than that of the MoS$_2$ FET under illumination at the same laser power (P$_{light}$). When the P$_{light}$ and V$_d$ are 2.7 μW and 5 V, the photoresponsivity (R$_{ph}$=I$_{ph}$/P$_{light}$) of graphene-MoS$_2$ FET and MoS$_2$ FET are 67 mA/W and 1.5 mA/W, respectively, indicating that the graphene-MoS$_2$ FET has a much better photoresponse than does the MoS$_2$ FET. As the light is turned on or switched off, a sharp increase or decrease in I$_{ph}$ is observed, respectively. The ON-OFF photoswitching behaviors of the graphene-MoS$_2$ FET and MoS$_2$ FET are shown in Figure 9b-d, which are well retained after three photoillumination cycles at different P$_{light}$.

Our aforementioned results indicate that graphene-MoS$_2$ FET devices show better performance such as much higher ON/OFF ratio and better photoresponsivity than the FET devices based on MoS$_2$ nanosheets, which is consistent with previous reports.[32, 35-36] Other heterostructures of 2D nanosheets, such as MoS$_2$-WSe$_2$, MoS$_2$-TaS$_2$ and graphene-WSe$_2$, have also been prepared (Figure S11 in SI). The facile fabrication of these heterostructures offers the possibility to create new kinds of complex structures with unusual properties and functions.

**CONCLUSIONS**

In summary, a universal, rapid method is developed to transfer various low-dimensional nanomaterials, such as nanoparticles, nanowires and 2D nanosheets, onto diverse substrates with high fidelity. Many kinds of polymers, such as PMMA, PLLA and PLC, can be used as the carrier, which can be considerably removed by being dissolved in DCM at 50 ºC without thermal annealing. The whole transfer process is very fast, which can be finished within several minutes. Essentially, cracks and wrinkles are avoided in the transferred 2D nanosheets due to the dry transfer process. As a proof of concept, various kinds of vertical heterostructures of 2D nanosheets have been successfully prepared. The graphene-MoS$_2$ FET shows not only higher ON/OFF ratio but also much better photoresponse compared to the MoS$_2$ FET. We believe that our method will greatly facilitate the fabrication of various multi-dimensional and functional architectures from low-dimensional nanomaterials, which could be used for various applications.

**Methods**

**Materials.** Poly(L-lactic acid) (PLLA, intrinsic viscosity (IV): 2.38) was purchased from Bio Invigor Corporation, Taiwan. 70/30 L-lactide-ε-caprolactone copolymer (PLC) (Mw = 202,000 g/mol) were purchased from Purac Biomaterials. Dichloromethane (DCM) and poly(methylmethacrylate) (PMMA, Mw = 996,000 g/mol) were purchased from Sigma-Aldrich

Pte Ltd, Singapore. Natural graphite was purchased from NGS Naturgraphit GmbH, Germany. MoS$_2$ crystal was purchased from SPI Supplies, USA. Single crystals of WSe$_2$ and TaS$_2$ were purchased from Nanoscience Instruments Inc., USA. All chemicals were used as received without further purification.

**Transfer of low-dimensional nanomaterials.** The scheme, corresponding photos and videos of peeling off, transfer and clean processes are shown in Figure 1, Figure S12 in SI, and videos S1 to S3. The detailed procedure is described as follows. Low-dimensional nanomaterials (*e.g.* nanoparticles (NPs), nanowires (NWs), and graphene or transition metal dichalcogenide (TMD) nanosheets) were firstly deposited on a freshly cleaned 90 nm SiO$_2$/Si substrate by using mechanical exfoliation method or spin-coating (Figure 1a and Figure S12a in SI). Polymer (PMMA, PLLA or PLC) in dichloromethane (DCM) solution (3.0 wt. %) was spin-coated on the aforementioned nanomaterials deposited on the 90 nm SiO$_2$/Si substrate at 3,000 rpm (Figure 1b and Figure S12b in SI), followed by heating on a hotplate at 50 $^\circ$C for 3 min to remove DCM. About 1-mm-wide polymer strips were then removed by a blade at the edges of the substrate to expose the hydrophilic SiO$_2$ surface (Figure 1c and Figure S12c in SI). After that, a 1 to 2-mm-thick PDMS film was brought into conformal contact with the polymer-coated substrate (Figure 1d and video S1). A small water droplet (~30 μL) was then dropped at one edge of the polymer-coated substrate (Figure 1e and video S1). Since water can easily penetrate the hydrophobic-hydrophilic interface at nanoscale, the hydrophobic polymer film and hydrophilic SiO$_2$/Si substrate can be successfully separated within several seconds with the aid of water (video S1). Thus the nanomaterials, which adhered to the carrier polymer coated by the PDMS film as a rigid support to avoid the mechanical stress, were detached from the original substrate (Figure 1f and Figure S12d in SI). Due to the hydrophobicity of the carrier polymer and PDMS, water droplet, which was condensed at the corner of the polymer film, could be easily removed. The PDMS/polymer/nanomaterial film was then brought into contact with the target substrate and heated on a hotplate at 50 °C (Figure 1g and video S2). Because of the weaker adhesion between the PDMS film and carrier polymer compared to that between the carrier polymer and target substrate, the PDMS film was easily peeled off from the carrier polymer which adhered to the target substrate (Figure 1h, video S2, and Figure S12e,f in SI). The carrier polymer was then completely removed by being dissolved in DCM at 50 °C within several tens of seconds (video S3). Finally, the nanomaterials with clean surface were left on the target substrate (Figure 1i and Figure S12g,h in SI).

Gold nanoparticles (Au NPs), pre-adsorbed on the aminosilane-patterned 90 nm $SiO_2$/Si substrate[20] were transferred by using the aforementioned procedure. Silver nanowires (Ag NWs) were spin-coated on 90 nm $SiO_2$/Si substrate followed by the deposition of graphene oxide (GO) nanosheets to form the Ag NW-GO composite, which was transferred from the 90 nm $SiO_2$/Si substrate onto 300 nm $SiO_2$/Si by using the aforementioned procedure.

**Atomic force microscopy (AFM) and optical microscopy (OM) characterization.** AFM (Dimension ICON with NanoScope V controller, Bruker, USA) and optical microscope (Eclipse LV100D with a 100×, 0.9 numerical aperture (NA) objective, Nikon) were used to characterize the low-dimensional nanomaterials before and after transfer. OM was also used to determine the thickness of 2D nanosheets on 90 nm $SiO_2$/Si.[40] Fabrication of the heterostructures of 2D nanosheets was performed by using our transfer method with a micromanipulator under the optical microscope.

**Raman Spectroscopy.** Analysis of the single- and few-layer graphene nanosheets by Raman spectroscopy was carried out on a WITec CRM200 confocal Raman microscopy system with the excitation line of 488 nm and an air cooling charge coupled device (CCD) as the detector (WITec Instruments Corp, Germany). Analysis of the single- and few-layer $MoS_2$ and $WSe_2$ nanosheets by Raman spectroscopy was carried out on a Renishaw inVia Raman microscope with laser light ($\lambda$= 532 nm).[41] The Raman band of a silicon wafer at 520 cm$^{-1}$ was used as a reference to calibrate the spectrometer.

**Device fabrication and characterization.** The source and drain electrodes of FET devices were fabricated by photolithography or e-beam lithography. The 5 nm Cr/50 nm Au source-drain electrodes were deposited by using thermal evaporator. After removal of photoresist, the electrical properties of FETs were tested by the Keithley 4200 semiconductor characterization system in air at room temperature. Photoresponse test was performed by using a 532 nm laser.

**Acknowledgement.**

This work was supported by MOE under AcRF Tier 2 (ARC 26/13, No. MOE2013-T2-1-034), AcRF Tier 1 (RG 61/12, RGT18/13 and RG5/13), and Start-Up Grant (M4080865.070.706022) in Singapore. This Research is also conducted by NTU-HUJ-BGU Nanomaterials for Energy and Water Management Programme under the Campus for Research Excellence and Technological Enterprise (CREATE), that is supported by the National Research Foundation, Prime Minister's Office, Singapore.

**Corresponding Author**

*E-mail: hzhang@ntu.edu.sg, hzhang166@yahoo.com.

Phone: +65-6790-5175. Fax: +65-6790-9081

Website: http://www.ntu.edu.sg/home/hzhang/

*Conflict of Interest:* The authors declare no competing financial interest.

*Supporting Information Available.* Raman spectra of grapheme, MoS$_2$, and WSe$_2$ nanosheets before and after transfer. $I_d$-$V_g$ curves of FETs based on pristine and transferred MoS$_2$ nanosheets. Clean transfer of 2D nanosheets by using PMMA and PLC as carrier polymer. Transfer of hydrophilic GO and mica sheets. Transfer of MoS$_2$ nanosheets onto various substrates. Transfer of Au NP-graphene composite from 90 nm to 300 nm SiO$_2$/Si substrate. Fabrication of MoS$_2$-WSe$_2$, MoS$_2$-TaS$_2$, and graphene-WSe$_2$ heterostructures. Photographs of the transfer and clean processes. These materials are available free of charge *via* the Internet at http://pubs.acs.org.

**REFERENCES AND NOTES**

1. Carlson, A.; Bowen, A. M.; Huang, Y.; Nuzzo, R. G.; Rogers, J. A. Transfer Printing Techniques for Materials Assembly and Micro/Nanodevice Fabrication. *Adv. Mater.* **2012,** *24*, 5284-5318.
2. Meitl, M. A.; Zhu, Z. T.; Kumar, V.; Lee, K. J.; Feng, X.; Huang, Y. Y.; Adesida, I.; Nuzzo, R. G.; Rogers, J. A. Transfer Printing by Kinetic Control of Adhesion to an Elastomeric Stamp. *Nat. Mater.* **2006,** *5*, 33-38.
3. Li, X.; Cai, W.; An, J.; Kim, S.; Nah, J.; Yang, D.; Piner, R.; Velamakanni, A.; Jung, I.; Tutuc, E*.; et al.* Large-Area Synthesis of High-Quality and Uniform Graphene Films on Copper Foils. *Science* **2009,** *324*, 1312-1314.
4. Jiao, L.; Fan, B.; Xian, X.; Wu, Z.; Zhang, J.; Liu, Z. Creation of Nanostructures with Poly(Methyl Methacrylate)-Mediated Nanotransfer Printing. *J. Am. Chem. Soc.* **2008,** *130*, 12612-12613.
5. Liu, Y.; Cheng, R.; Liao, L.; Zhou, H.; Bai, J.; Liu, G.; Liu, L.; Huang, Y.; Duan, X. Plasmon Resonance Enhanced Multicolour Photodetection by Graphene. *Nat. Commun.* **2011,** *2*, 579.
6. Cheng, R.; Bai, J.; Liao, L.; Zhou, H.; Chen, Y.; Liu, L.; Lin, Y. C.; Jiang, S.; Huang, Y.; Duan, X. High-Frequency Self-Aligned Graphene Transistors with Transferred Gate Stacks. *Proc. Natl. Acad. Sci. U.S.A.* **2012,** *109*, 11588-11592.
7. Schneider, G. F.; Calado, V. E.; Zandbergen, H.; Vandersypen, L. M.; Dekker, C. Wedging Transfer of Nanostructures. *Nano Lett.* **2010,** *10*, 1912-1916.
8. Lee, C. H.; Kim, D. R.; Cho, I. S.; William, N.; Wang, Q.; Zheng, X. Peel-and-Stick: Fabricating Thin Film Solar Cell on Universal Substrates. *Sci. Rep.* **2012,** *2*, 1000.
9. Lin, Y.-C.; Lu, C.-C.; Yeh, C.-H.; Jin, C.; Suenaga, K.; Chiu, P.-W. Graphene Annealing: How Clean Can It Be? *Nano Lett.* **2011,** *12*, 414-419.
10. Suk, J. W.; Kitt, A.; Magnuson, C. W.; Hao, Y.; Ahmed, S.; An, J.; Swan, A. K.; Goldberg, B. B.; Ruoff, R. S. Transfer of CVD-Grown Monolayer Graphene onto Arbitrary Substrates. *ACS Nano* **2011,** *5*, 6916-6924.
11. Suk, J. W.; Lee, W. H.; Lee, J.; Chou, H.; Piner, R. D.; Hao, Y.; Akinwande, D.; Ruoff, R. S. Enhancement of the Electrical Properties of Graphene Grown by Chemical Vapor Deposition *via* Controlling the Effects of Polymer Residue. *Nano Lett.* **2013,** *13*, 1462-1467.
12. Li, X.; Zhu, Y.; Cai, W.; Borysiak, M.; Han, B.; Chen, D.; Piner, R. D.; Colombo, L.; Ruoff, R. S. Transfer of Large-Area Graphene Films for High-Performance Transparent Conductive Electrodes. *Nano Lett.* **2009,** *9*, 4359-4363.
13. Reina, A.; Son, H.; Jiao, L.; Fan, B.; Dresselhaus, M. S.; Liu, Z.; Kong, J. Transferring and Identification of Single- and Few-Layer Graphene on Arbitrary Substrates. *J. Phys. Chem. C* **2008,** *112*, 17741-17744.

14. Liang, X.; Sperling, B. A.; Calizo, I.; Cheng, G.; Hacker, C. A.; Zhang, Q.; Obeng, Y.; Yan, K.; Peng, H.; Li, Q.*; et al.* Toward Clean and Crackless Transfer of Graphene. *ACS Nano* **2011,** *5*, 9144-9153.
15. Banerjee, S.; Shim, J.; Rivera, J.; Jin, X.; Estrada, D.; Solovyeva, V.; You, X.; Pak, J.; Pop, E.; Aluru, N.*; et al.* Electrochemistry at the Edge of a Single Graphene Layer in a Nanopore. *ACS Nano* **2012,** *7*, 834-843.
16. Pu, J.; Yomogida, Y.; Liu, K. K.; Li, L. J.; Iwasa, Y.; Takenobu, T. Highly Flexible $MoS_2$ Thin-Film Transistors with Ion Gel Dielectrics. *Nano Lett.* **2012,** *12*, 4013-4017.
17. van der Zande, A. M.; Huang, P. Y.; Chenet, D. A.; Berkelbach, T. C.; You, Y.; Lee, G. H.; Heinz, T. F.; Reichman, D. R.; Muller, D. A.; Hone, J. C. Grains and Grain Boundaries in Highly Crystalline Monolayer Molybdenum Disulphide. *Nat. Mater.* **2013,** *12*, 554-561.
18. Bae, S.; Kim, H.; Lee, Y.; Xu, X.; Park, J. S.; Zheng, Y.; Balakrishnan, J.; Lei, T.; Kim, H. R.; Song, Y. I.*; et al.* Roll-to-Roll Production of 30-Inch Graphene Films for Transparent Electrodes. *Nat. Nanotechnol.* **2010,** *5*, 574-578.
19. Song, J.; Kam, F. Y.; Png, R. Q.; Seah, W. L.; Zhuo, J. M.; Lim, G. K.; Ho, P. K.; Chua, L. L. A General Method for Transferring Graphene onto Soft Surfaces. *Nat. Nanotechnol.* **2013,** *8*, 356-362.
20. Li, H.; Zhang, J.; Zhou, X. Z.; Lu, G.; Yin, Z. Y.; Li, G. P.; Wu, T.; Boey, F.; Venkatraman, S. S.; Zhang, H. Aminosilane Micropatterns on Hydroxyl-Terminated Substrates: Fabrication and Applications. *Langmuir* **2010,** *26*, 5603-5609.
21. Salvatore, G. A.; Münzenrieder, N.; Barraud, C.; Petti, L.; Zysset, C.; Büthe, L.; Ensslin, K.; Tröster, G. Fabrication and Transfer of Flexible Few-Layers $MoS_2$ Thin Film Transistors to Any Arbitrary Substrate. *ACS Nano* **2013,** *7*, 8809-8815.
22. Lee, Y. H.; Yu, L.; Wang, H.; Fang, W.; Ling, X.; Shi, Y.; Lin, C. T.; Huang, J. K.; Chang, M. T.; Chang, C. S.*; et al.* Synthesis and Transfer of Single-Layer Transition Metal Disulfides on Diverse Surfaces. *Nano Lett.* **2013,** *13*, 1852-1857.
23. Li, H.; Yin, Z. Y.; He, Q. Y.; Huang, X.; Lu, G.; Fam, D. W. H.; Tok, A. I. Y.; Zhang, Q.; Zhang, H. Fabrication of Single- and Multilayer $MoS_2$ Film-Based Field-Effect Transistors for Sensing No at Room Temperature. *Small* **2012,** *8*, 63-67.
24. Yin, Z. Y.; Li, H.; Li, H.; Jiang, L.; Shi, Y. M.; Sun, Y. H.; Lu, G.; Zhang, Q.; Chen, X. D.; Zhang, H. Single-Layer $MoS_2$ Phototransistors. *ACS Nano* **2012,** *6*, 74-80.
25. Huang, X.; Zeng, Z.; Zhang, H. Metal Dichalcogenide Nanosheets: Preparation, Properties and Applications. *Chem. Soc. Rev.* **2013,** *42*, 1934-1946.
26. Zhan, Y.; Liu, Z.; Najmaei, S.; Ajayan, P. M.; Lou, J. Large-Area Vapor-Phase Growth and Characterization of $MoS_2$ Atomic Layers on a $SiO_2$ Substrate. *Small* **2012,** *8*, 966-971.
27. Cheng, Z.; Zhou, Q.; Wang, C.; Li, Q.; Wang, C.; Fang, Y. Toward Intrinsic Graphene Surfaces: A Systematic Study on Thermal Annealing and Wet-Chemical Treatment of $SiO_2$-Supported Graphene Devices. *Nano Lett.* **2011,** *11*, 767-771.
28. Dean, C. R.; Young, A. F.; Meric, I.; Lee, C.; Wang, L.; Sorgenfrei, S.; Watanabe, K.; Taniguchi, T.; Kim, P.; Shepard, K. L.*; et al.* Boron Nitride Substrates for High-Quality Graphene Electronics. *Nat. Nanotechnol.* **2010,** *5*, 722-726.
29. Britnell, L.; Gorbachev, R. V.; Jalil, R.; Belle, B. D.; Schedin, F.; Mishchenko, A.; Georgiou, T.; Katsnelson, M. I.; Eaves, L.; Morozov, S. V.*; et al.* Field-Effect Tunneling Transistor Based on Vertical Graphene Heterostructures. *Science* **2012,** *335*, 947-950.
30. Bertolazzi, S.; Krasnozhon, D.; Kis, A. Nonvolatile Memory Cells Based on $MoS_2$/Graphene Heterostructures. *ACS Nano* **2013,** *7*, 3246-3252.
31. Chan, M. Y.; Komatsu, K.; Li, S. L.; Xu, Y.; Darmawan, P.; Kuramochi, H.; Nakaharai, S.; Aparecido-Ferreira, A.; Watanabe, K.; Taniguchi, T.*; et al.* Suppression of Thermally Activated Carrier Transport in Atomically Thin $MoS_2$ on Crystalline Hexagonal Boron Nitride Substrates. *Nanoscale* **2013,** *5*, 9572-9576.

32. Choi, M. S.; Lee, G. H.; Yu, Y. J.; Lee, D. Y.; Lee, S. H.; Kim, P.; Hone, J.; Yoo, W. J. Controlled Charge Trapping by Molybdenum Disulphide and Graphene in Ultrathin Heterostructured Memory Devices. *Nat. Commun.* **2013,** *4*, 1624.
33. Geim, A. K.; Grigorieva, I. V. Van Der Waals Heterostructures. *Nature* **2013,** *499*, 419-425.
34. Georgiou, T.; Jalil, R.; Belle, B. D.; Britnell, L.; Gorbachev, R. V.; Morozov, S. V.; Kim, Y. J.; Gholinia, A.; Haigh, S. J.; Makarovsky, O.*; et al.* Vertical Field-Effect Transistor Based on Graphene-WS$_2$ Heterostructures for Flexible and Transparent Electronics. *Nat. Nanotechnol.* **2013,** *8*, 100-103.
35. Roy, K.; Padmanabhan, M.; Goswami, S.; Sai, T. P.; Ramalingam, G.; Raghavan, S.; Ghosh, A. Graphene-MoS$_2$ Hybrid Structures for Multifunctional Photoresponsive Memory Devices. *Nat. Nanotechnol.* **2013,** *8*, 826-830.
36. Yu, W. J.; Li, Z.; Zhou, H.; Chen, Y.; Wang, Y.; Huang, Y.; Duan, X. Vertically Stacked Multi-Heterostructures of Layered Materials for Logic Transistors and Complementary Inverters. *Nat. Mater.* **2013,** *12*, 246-252.
37. Yu, W. J.; Liu, Y.; Zhou, H.; Yin, A.; Li, Z.; Huang, Y.; Duan, X. Highly Efficient Gate-Tunable Photocurrent Generation in Vertical Heterostructures of Layered Materials. *Nat. Nanotechnol.* **2013,** *8*, 952-958.
38. Lee, H. S.; Min, S. W.; Chang, Y. G.; Park, M. K.; Nam, T.; Kim, H.; Kim, J. H.; Ryu, S.; Im, S. MoS$_2$ Nanosheet Phototransistors with Thickness-Modulated Optical Energy Gap. *Nano Lett.* **2012,** *12*, 3695-3700.
39. Lopez-Sanchez, O.; Lembke, D.; Kayci, M.; Radenovic, A.; Kis, A. Ultrasensitive Photodetectors Based on Monolayer MoS$_2$. *Nat. Nanotechnol.* **2013,** *8*, 497-501.
40. Li, H.; Wu, J.; Huang, X.; Lu, G.; Yang, J.; Lu, X.; Xiong, Q.; Zhang, H. Rapid and Reliable Thickness Identification of Two-Dimensional Nanosheets Using Optical Microscopy. *ACS Nano* **2013,** *7*, 10344-10353.
41. Li, H.; Lu, G.; Wang, Y.; Yin, Z. Y.; Cong, C.; He, Q.; Wang, L.; Ding, F.; Yu, T.; Zhang, H. Mechanical Exfoliation and Characterization of Single- and Few-Layer Nanosheets of WSe$_2$, TaS$_2$, and TaSe$_2$. *Small* **2013,** *9*, 1974-1981.

**Figure Caption**

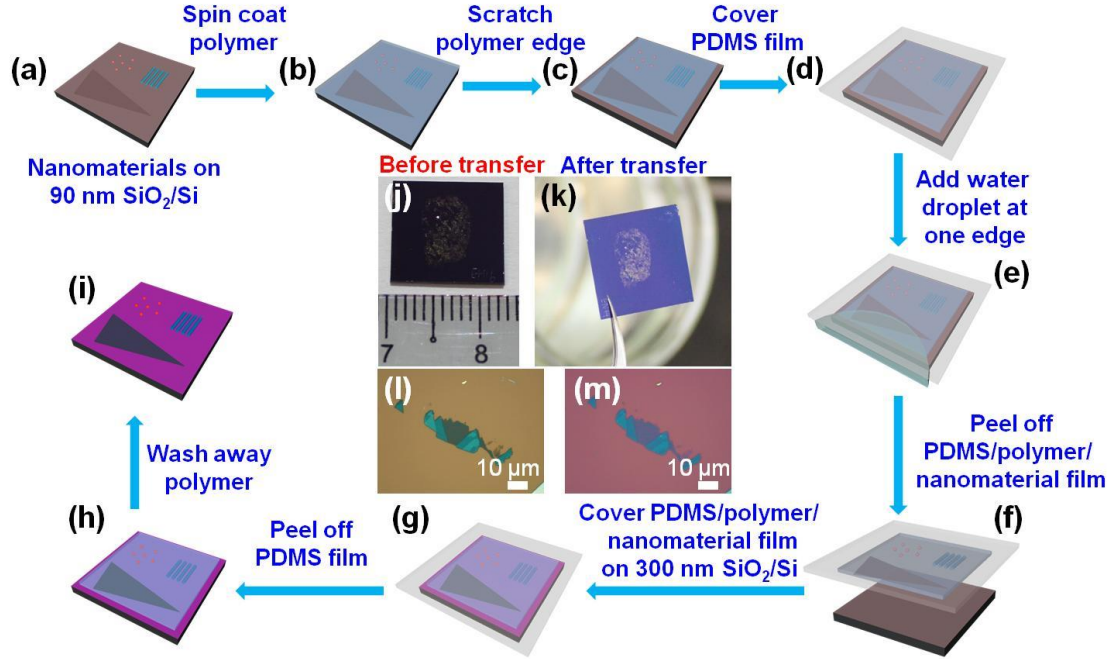

**Figure 1.** Schematic illustration of high-fidelity transfer of nanostructures. (a) Low-dimensional nanomaterials (*e.g.* nanoparticle, nanowires, and nanosheets) are deposited on a 90 nm $SiO_2$/Si substrate. (b) A polymer film as a carrier is spin-coated onto the nanomaterials deposited on $SiO_2$/Si substrate. (c) 1-mm-wide polymer strips are removed at the edges of polymer-coated $SiO_2$/Si substrate to expose the hydrophilic $SiO_2$ surface. (d) A 1 to 2-mm-thick PDMS film is brought into conformal contact with the carrier polymer. (e) A drop of water is deposited at one edge of $SiO_2$/Si substrate to separate the PDMS/polymer/nanomaterial film from $SiO_2$/Si substrate. (f) The PDMS/polymer/nanomaterial film is peeled off from $SiO_2$/Si substrate. (g) The PDMS/polymer/nanomaterial film is brought into contact with a 300 nm $SiO_2$/Si substrate. (h) The PDMS film is removed from the carrier polymer film. (i) The carrier polymer is washed away after it is dissolved in DCM at 50 ºC. The nanomaterials are left on the target substrate. (j) Photograph of $MoS_2$ flakes deposited on a 90 nm $SiO_2$/Si substrate. (k) Photograph of $MoS_2$ flakes in (j) transferred onto a 300 nm $SiO_2$/Si substrate. (l and m) OM images of the $MoS_2$ nansoheets before (l) and after (m) transfer.

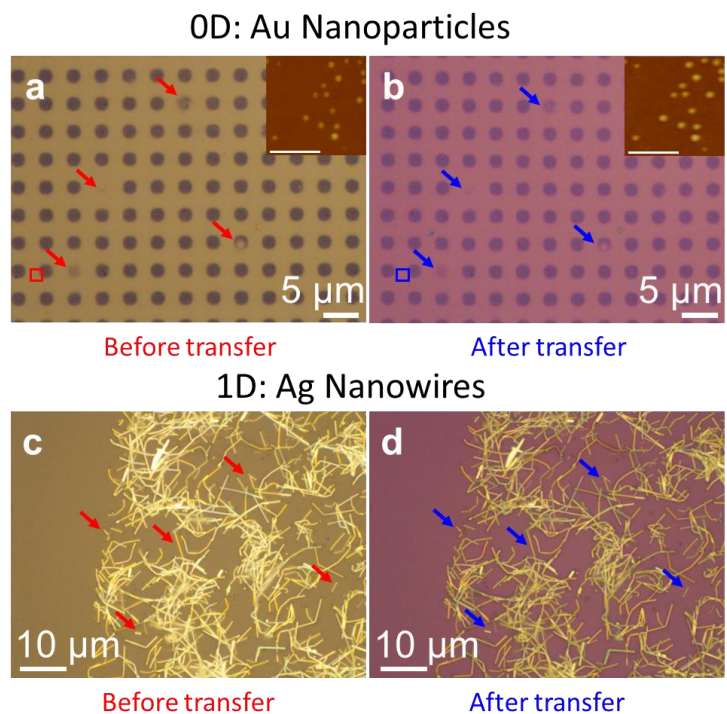

**Figure 2.** OM images of Au NPs array before (a) and after (b) transfer. Insets: AFM images of Au NPs before (a) and after (b) transfer shown in the red (a) and blue (b) squares, respectively. The scale bars represent 300 nm. OM images of Ag NWs before (c) and after (d) transfer. The red (a and c) or blue (b and d) arrows point to the same positions in the array of Au NPs or Ag NWs before and after transfer, respectively, indicating the precise-fidelity transfer.

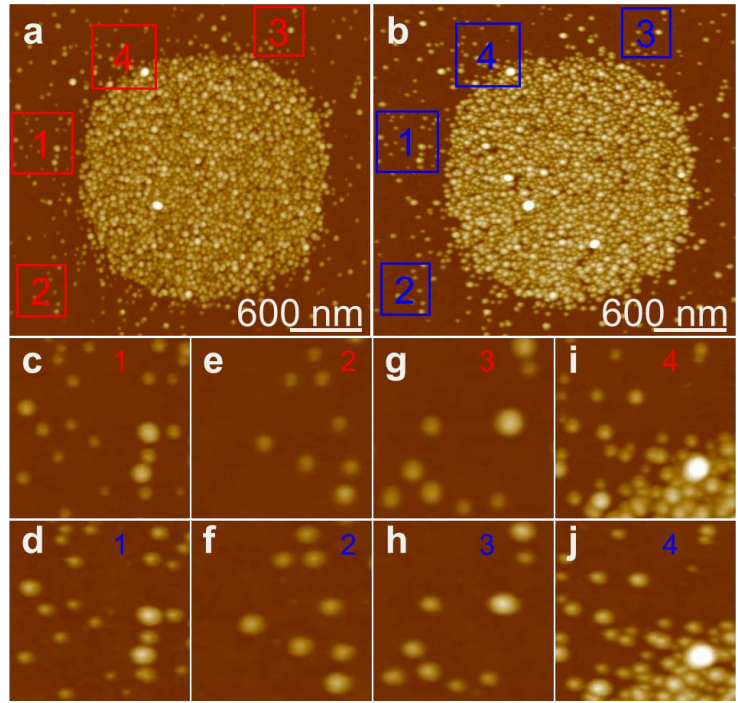

**Figure 3.** AFM images of Au NPs before (a) and after (b) transfer. (c to j) Magnified AFM images of the red and blue boxes labelled with 1, 2, 3 and 4 shown in (a) and (b), respectively. The AFM images indicate that individual Au NPs were transferred with high fidelity.

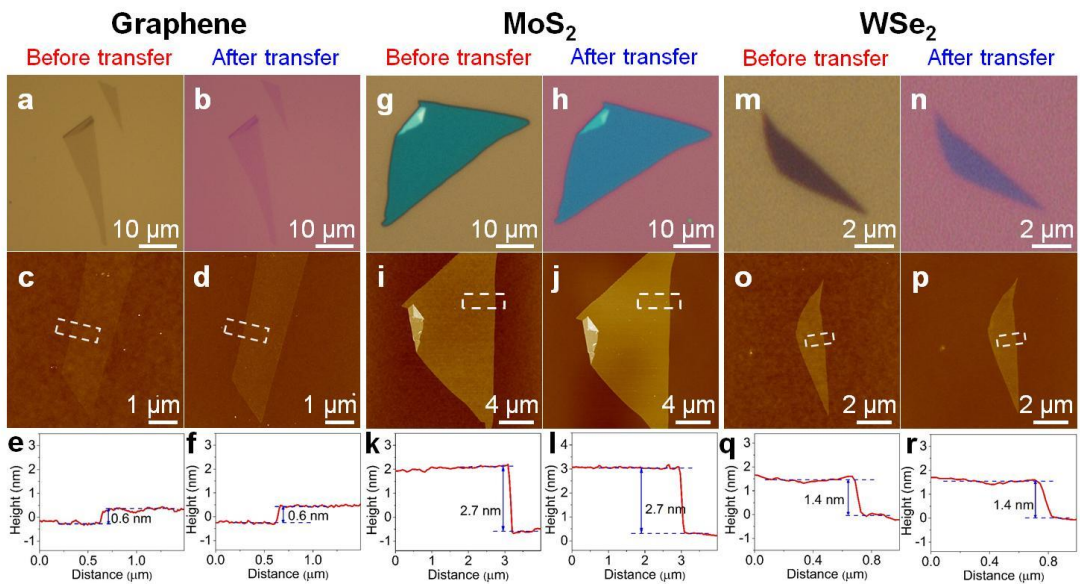

**Figure 4.** OM (a-b) and AFM (c-d) images of 1L graphene nanosheets before (a and c) and after (b and d) transfer. (e-f) The corresponding height profiles of the dashed rectangles shown in (c) and (d), respectively. OM (g-h) and AFM (i-j) images of 4L $MoS_2$ nanosheets before (g and i) and after (h and j) transfer. (k-l) The corresponding height profiles of the dashed rectangles shown in (i) and (j), respectively. OM (m-n) and AFM (o-p) images of 2L $WSe_2$ nanosheets before (m and o) and after (n and p) transfer. (q-r) The corresponding height profiles of the dashed rectangles shown in (o) and (p), respectively. The heights of 2D nanosheets remained unchanged after transfer, indicating no polymer residue was left on the transferred 2D nanosheets by using our high-fidelity transfer method.

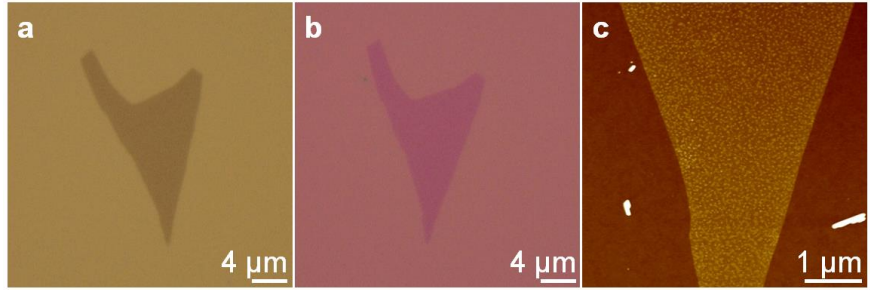

**Figure 5.** OM images of a graphene nanosheet transferred from (a) 90 nm $SiO_2$/Si onto (b) 300 nm $SiO_2$/Si by using the common PMMA-mediated transfer method.[13] (c) AFM image of graphene nanosheet transferred onto 300 nm $SiO_2$/Si after PMMA is removed by acetone. Many PMMA residues were left on the graphene nanosheet.

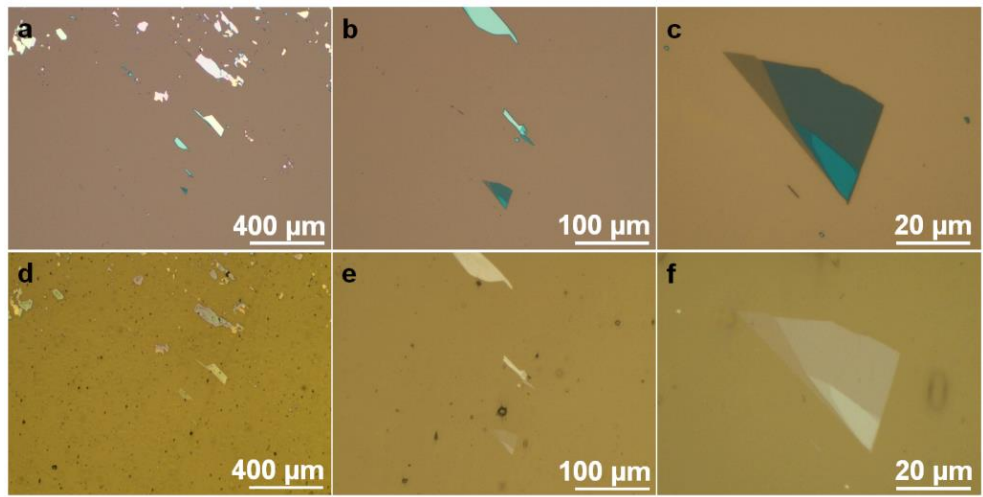

**Figure 6.** OM images of $MoS_2$ nanosheets transferred from (a-c) 90 nm $SiO_2$/Si onto (d-f) PET film.

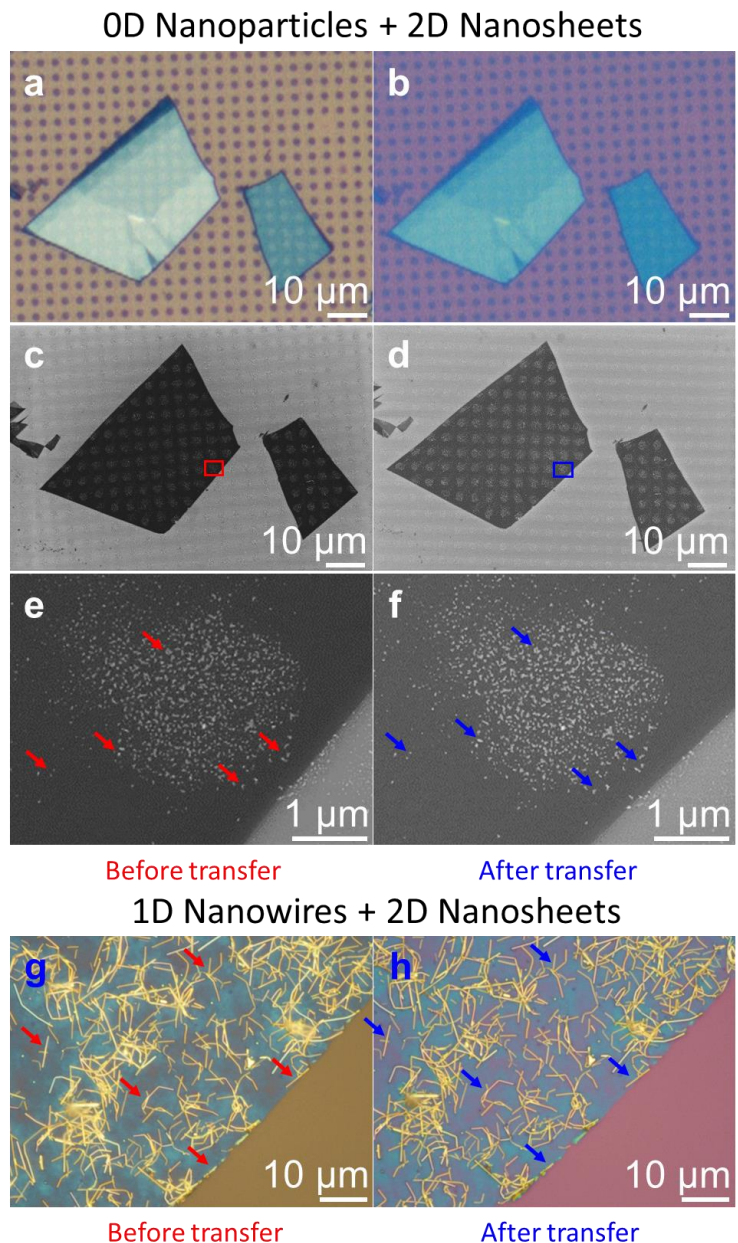

**Figure 7.** OM images of the Au NP-graphene composite and Ag NW-GO composite before (a and g) and after (b and h) transfer. (c-d) SEM images of Au NP-graphene composite shown in (a) and (b), respectively. (e-f) Magnified SEM images of red and blue rectangles shown in (c) and (d), respectively. The red (e and g) and blue (f and h) arrows show the features of Au NP-graphene and Ag NW-GO nanosheet before and after transfer, respectively, indicating the precise-fidelity transfer.

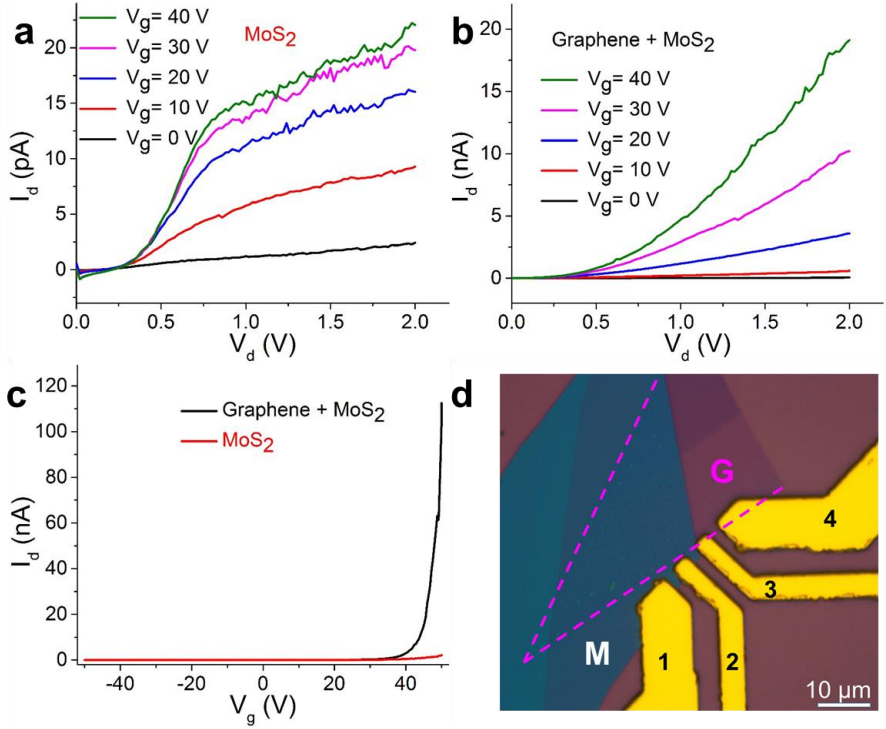

**Figure 8.** $I_d$-$V_d$ curves of (a) $MoS_2$ FET and (b) graphene-$MoS_2$ FET at $V_g$ ranging from 0 to 40 V. (c) $I_d$-$V_g$ curves of $MoS_2$ and graphene-$MoS_2$ FETs. (d) OM image of $MoS_2$ and graphene-$MoS_2$ FET. M and G indicate $MoS_2$ and graphene, respectively. The dashed line indicates the edge of the graphene.

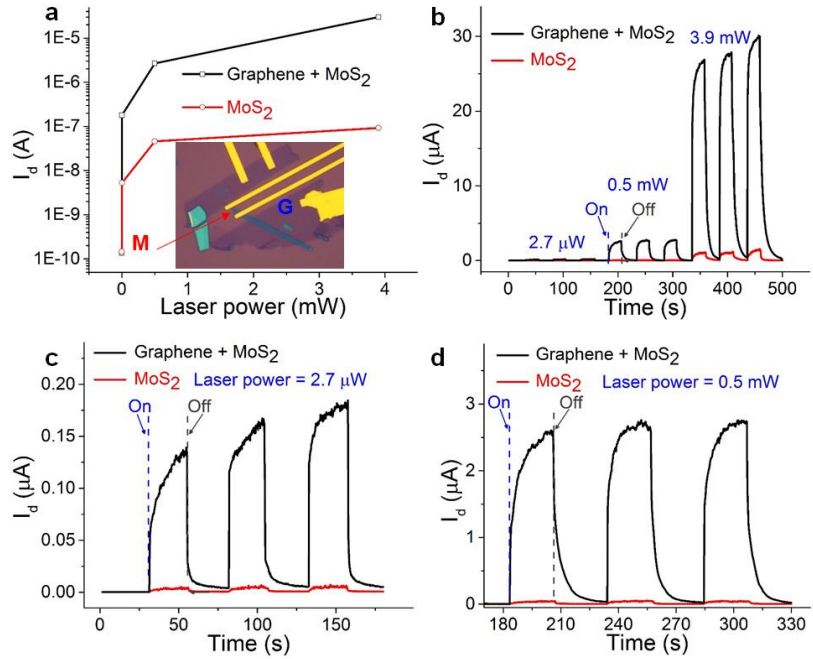

**Figure 9.** (a) Plots of $I_d$ *vs.* laser power of $MoS_2$ and graphene-$MoS_2$ FETs. Inset: OM image of the $MoS_2$ and graphene-$MoS_2$ FETs. M and G indicate $MoS_2$ and graphene, respectively. (b) Photoswitching behavior of the $MoS_2$ and graphene-$MoS_2$ FETs at laser power of 2.7 μW, 0.5 mW and 3.9 mW, respectively. (c) Photoswitching behavior of the $MoS_2$ and graphene-$MoS_2$ FETs at laser power of 2.7 μW. (d) Photoswitching behavior of the $MoS_2$ and graphene-$MoS_2$ FETs at laser power of 0.5 mW.

**Table 1.** Comparison of the root-mean-square (rms) surface roughness ($R_a$) of 2D nanosheets, measured based on $1 \times 1$ μm$^2$ AFM images, before and after transfer by using our clean transfer method and the common PMMA-mediated transfer method.

| Method | Material | Before transfer | After transfer |
| --- | --- | --- | --- |
| Using transfer method in this work (cleaning PLLA by DCM at 50 °C) | Graphene | 0.21 nm | 0.17 nm |
|  | MoS$_2$ | 0.19 nm | 0.18 nm |
|  | WSe$_2$ | 0.19 nm | 0.15 nm |
| Using common PMMA-mediated transfer method (cleaning PMMA by acetone) | Graphene |  | 0.49 nm[*] |
|  |  |  | 0.54 nm[**] |

[*]Obtained from the result shown in Figure 5c.

[**]Obtained from Ref. 25.

Table of contents graphic

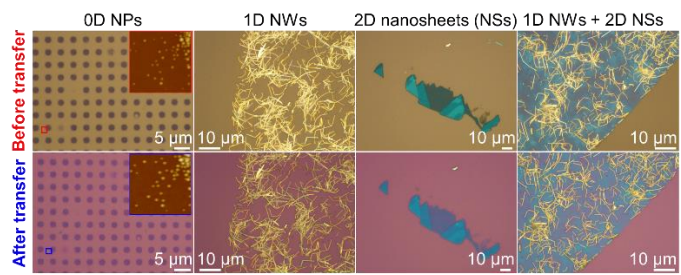

**Supporting Information**

# A Universal, Rapid Method for Clean Transfer of Nanostructures onto Various Substrates

*Hai Li,[1] Jumiati Wu,[1] Xiao Huang,[1] Zongyou Yin,[1] Juqing Liu,[1] Hua Zhang[1,*]*

[1]School of Materials Science and Engineering, Nanyang Technological University, 50 Nanyang Avenue, Singapore 639798, Singapore.

*To whom correspondence should be addressed.

Phone: +65-6790-5175. Fax: +65-6790-9081

E-mail: hzhang@ntu.edu.sg

Website: http://www.ntu.edu.sg/home/hzhang/

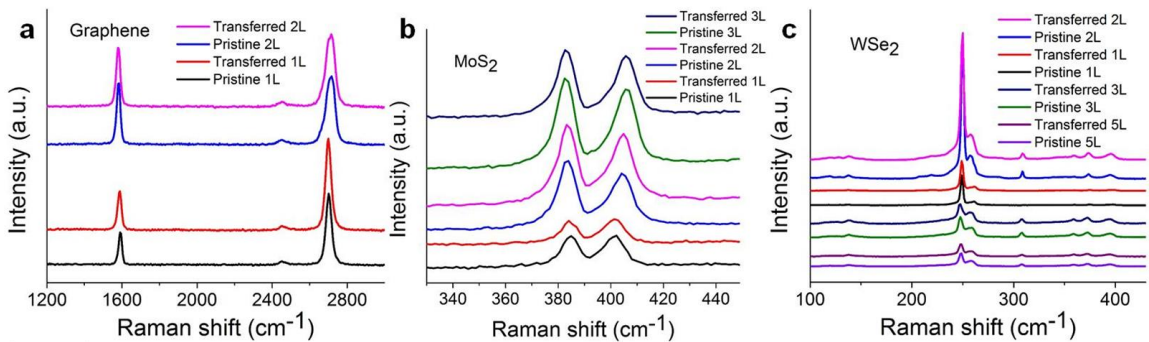

**Figure S1.** Raman spectra of (a) graphene, (b) $MoS_2$ and (c) $WSe_2$ nanosheets before and after transfer.

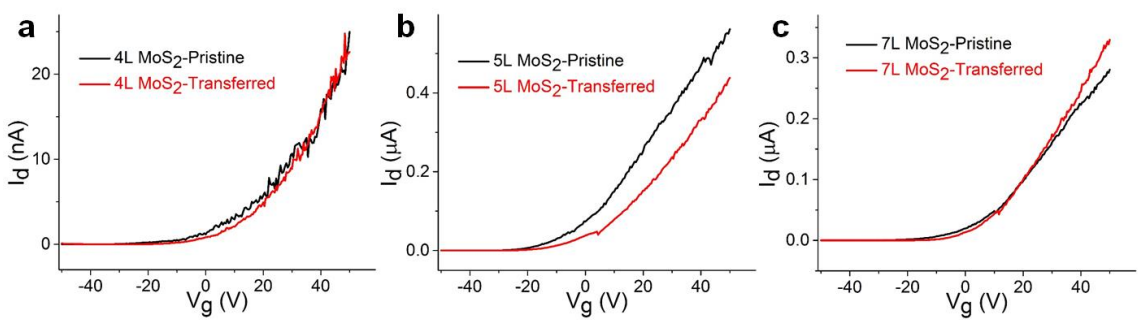

**Figure S2.** $I_d$-$V_g$ curves of FETs based on pristine and transferred (a) 4L, (b) 5L and (c) 7L $MoS_2$ nanosheets. The FETs based on the transferred $MoS_2$ nanosheets show the similar performance with the pristine $MoS_2$ nanosheets.

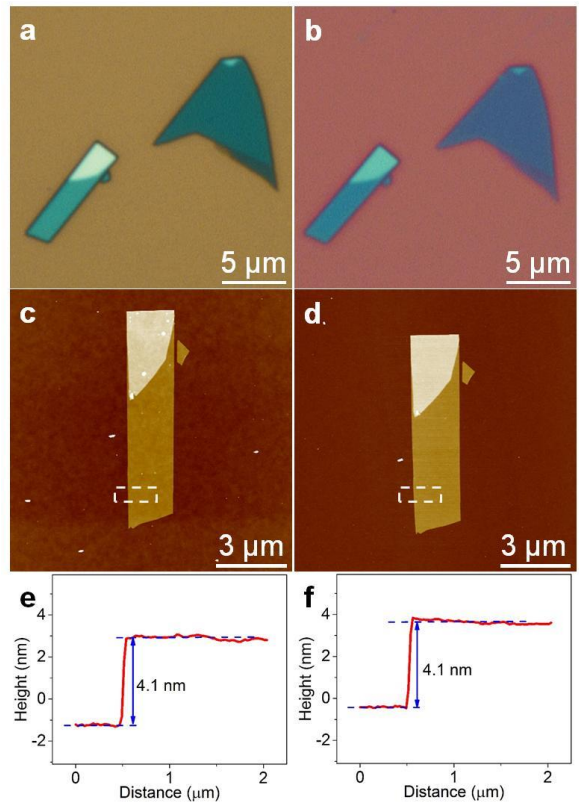

**Figure S3.** OM (a,b) and AFM (c,d) images of MoS$_2$ nanosheets before (a,c) and after (b,d) transfer by using PMMA as carrier polymer which was dissolved in DCM at 50 °C. (e,f) The corresponding height profiles of the dashed rectangles shown in (c) and (d), respectively. The height of the MoS$_2$ nanosheet remains unchanged after transfer, indicating PMMA can be considerably removed by DCM at 50 °C.

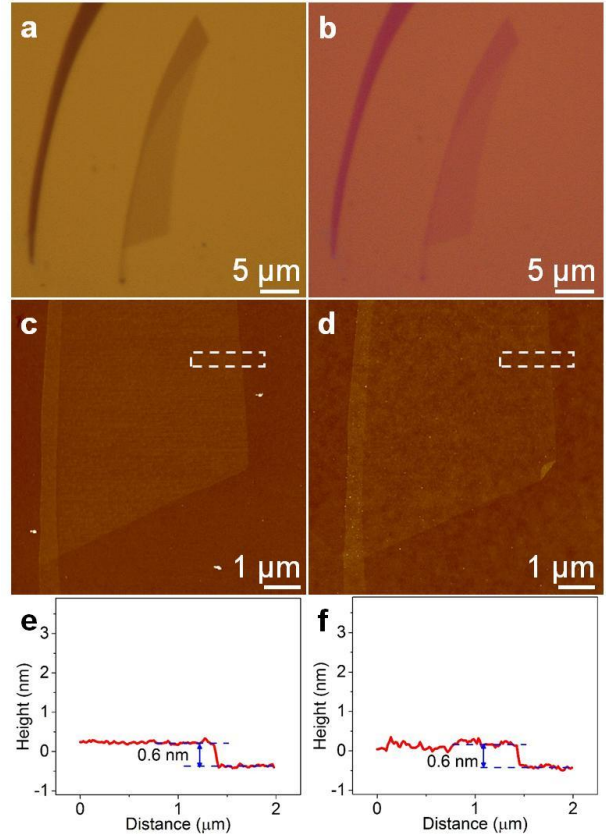

**Figure S4.** OM (a,b) and AFM (c,d) images of graphene nanosheets before (a,c) and after (b,d) transfer by using PLC as carrier polymer. The PLC film was removed after being dissolved in DCM at 50 °C. (e,f) The corresponding height profiles of the dashed rectangles shown in (c) and (d), respectively. The height of the graphene nanosheet remains unchanged after transfer, indicating PLC can be considerably removed by DCM at 50 °C.

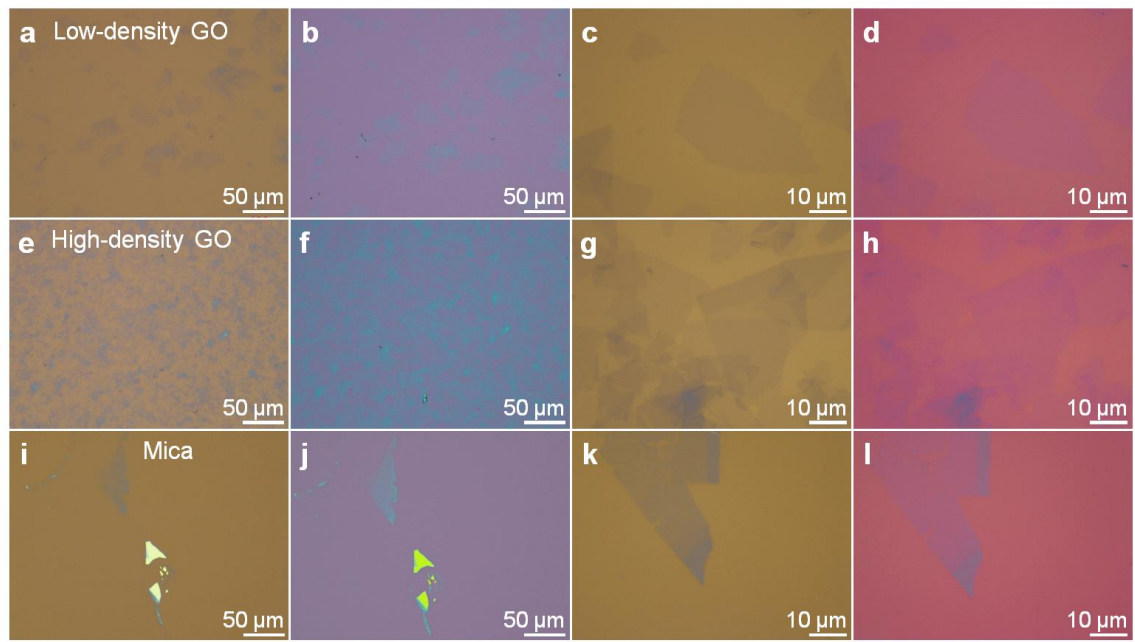

**Figure S5.** (a,b) OM images of low-density GO sheets on 90 nm $SiO_2$/Si substrate (a) and after being transferred onto 300 nm $SiO_2$/Si substrate (b). (c) and (d) show the magnified images of (a) and (b), respectively. (e,f) OM images of high-density GO sheets on 90 nm $SiO_2$/Si substrate (e) and after being transferred onto 300 nm $SiO_2$/Si substrate (f). (g) and (h) show the magnified images of (e) and (f), respectively. (i,j) OM images of mica sheets on 90 nm $SiO_2$/Si substrate (i) and after being transferred onto 300 nm $SiO_2$/Si substrate (j). (k) and (l) show the magnified images of (i) and (j), respectively.

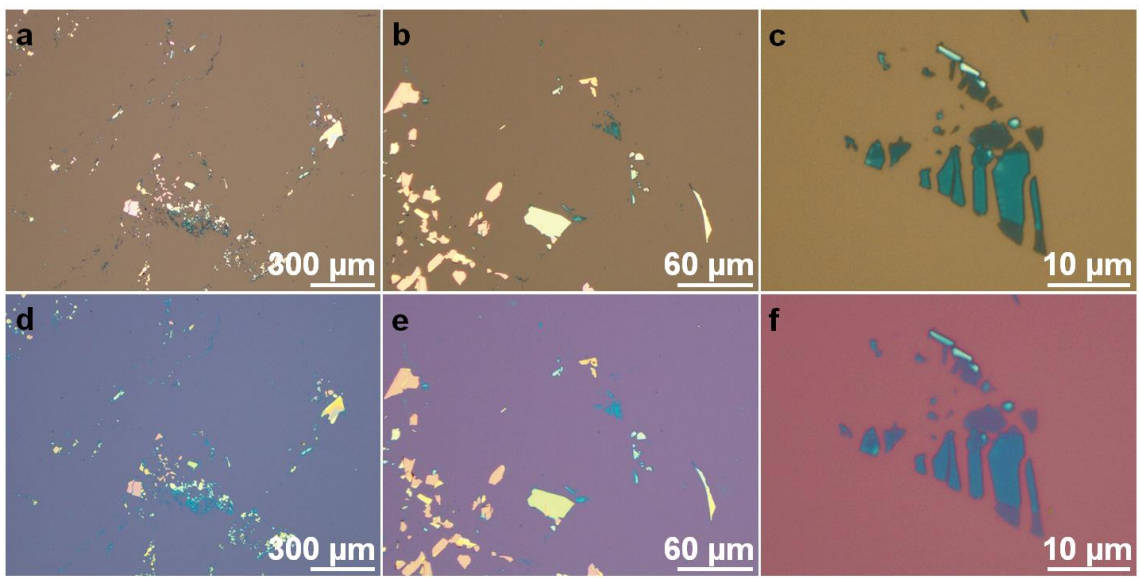

**Figure S6.** OM images of MoS$_2$ nanosheets transferred from (a-c) 90 nm SiO$_2$/Si onto (d-f) OTS-modified SiO$_2$/Si.[S1]

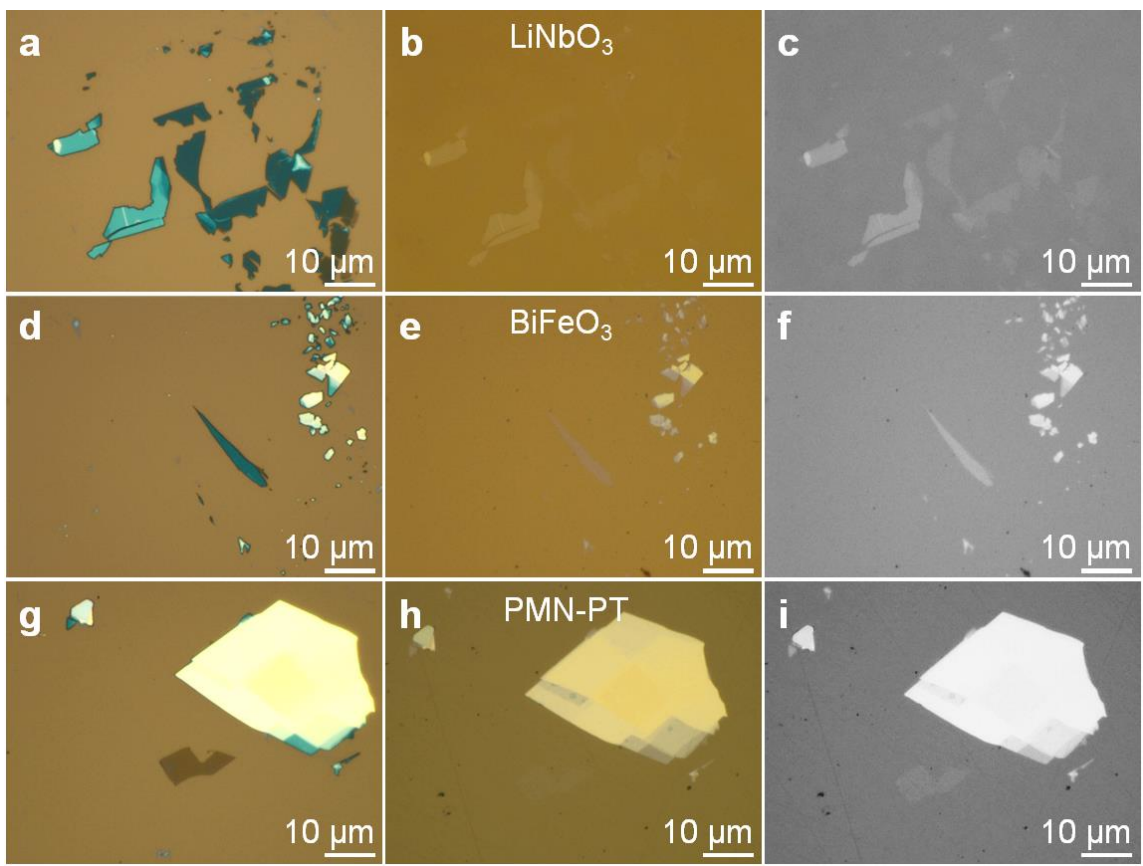

**Figure S7.** OM images of MoS$_2$ nanosheets transferred from (a,d,g) 90 nm SiO$_2$/Si onto (b) LiNbO$_3$, (e) BiFeO$_3$, and (h) PMN-PT substrates. (c,f,i) Gray scale images of (b), (e) and (h), respectively, which show better contrast for the transferred MoS$_2$ nanosheets.

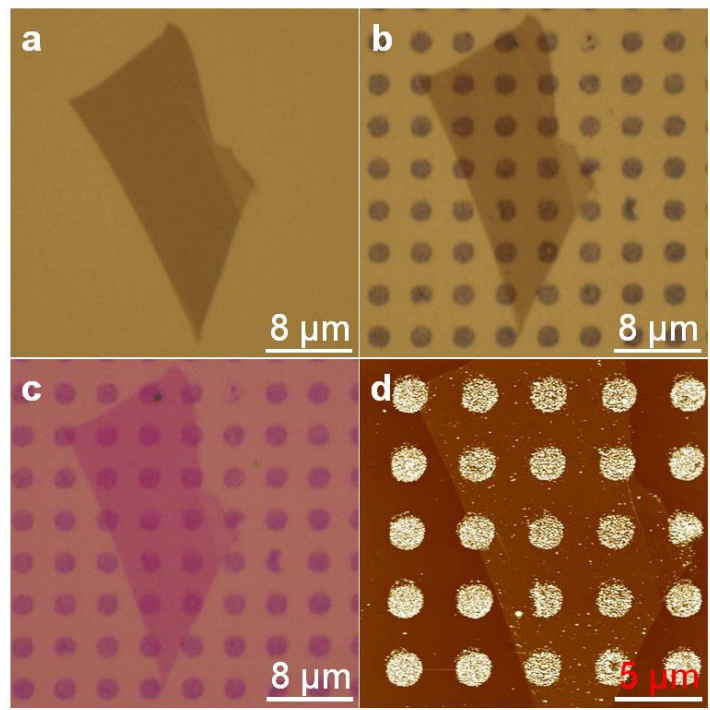

**Figure S8.** OM images of a graphene nanosheet on 90 nm $SiO_2$/Si substrate before (a) and after (b) Au NP array was transferred onto it. OM (c) and AFM (d) images of the Au NP-graphene composite transferred from 90 nm to 300 nm $SiO_2$/Si substrate.

Graphene nanosheet was firstly deposited on 90 nm $SiO_2$/Si by using the mechanical exfoliation method (Fig. S8a). Gold nanoparticles (Au NPs) adsorbed on aminosilane-patterned 90 nm $SiO_2$/Si substrate[S2] were transferred onto the graphene nanosheet to form the Au NP-graphene composite by using our clean transfer method (Fig. S8b). The composite was then transferred from the 90 nm $SiO_2$/Si substrate (Fig. S8b) onto 300 nm $SiO_2$/Si (Fig. S8c,d) by using our clean transfer method.

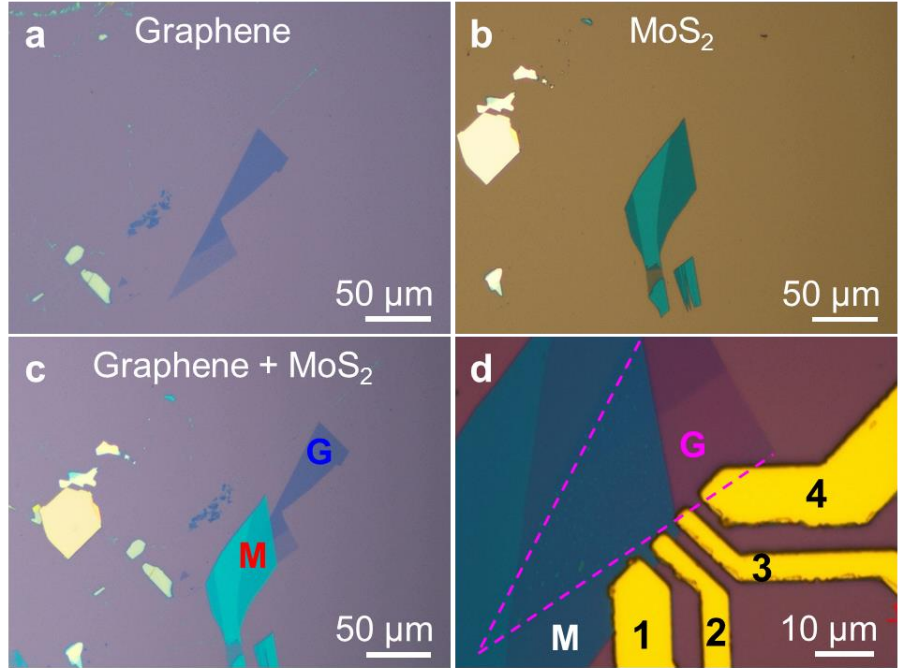

**Figure S9.** OM images of (a) graphene and (b) MoS$_2$ nanosheets. (c) OM image of a heterostructure formed by transferring the MoS$_2$ in (b) onto the graphene nanosheet in (a). (d) OM image of the FET based on the heterostructure of graphene and MoS$_2$. M and G indicate MoS$_2$ and graphene, respectively. The dashed line indicates the edge of the graphene.

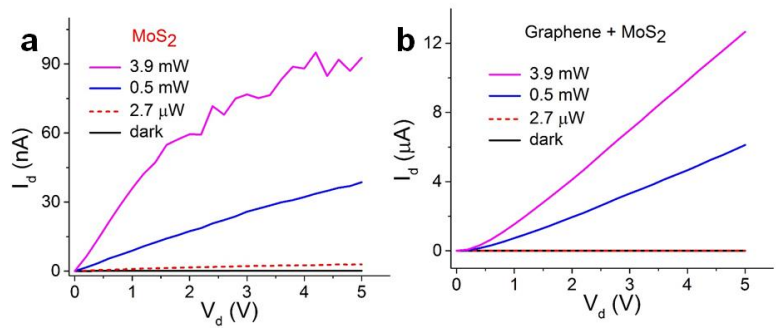

**Figure S10.** The source-drain current ($I_d$) curves of (a) MoS$_2$ FET and (b) graphene-MoS$_2$ FET at different laser power.

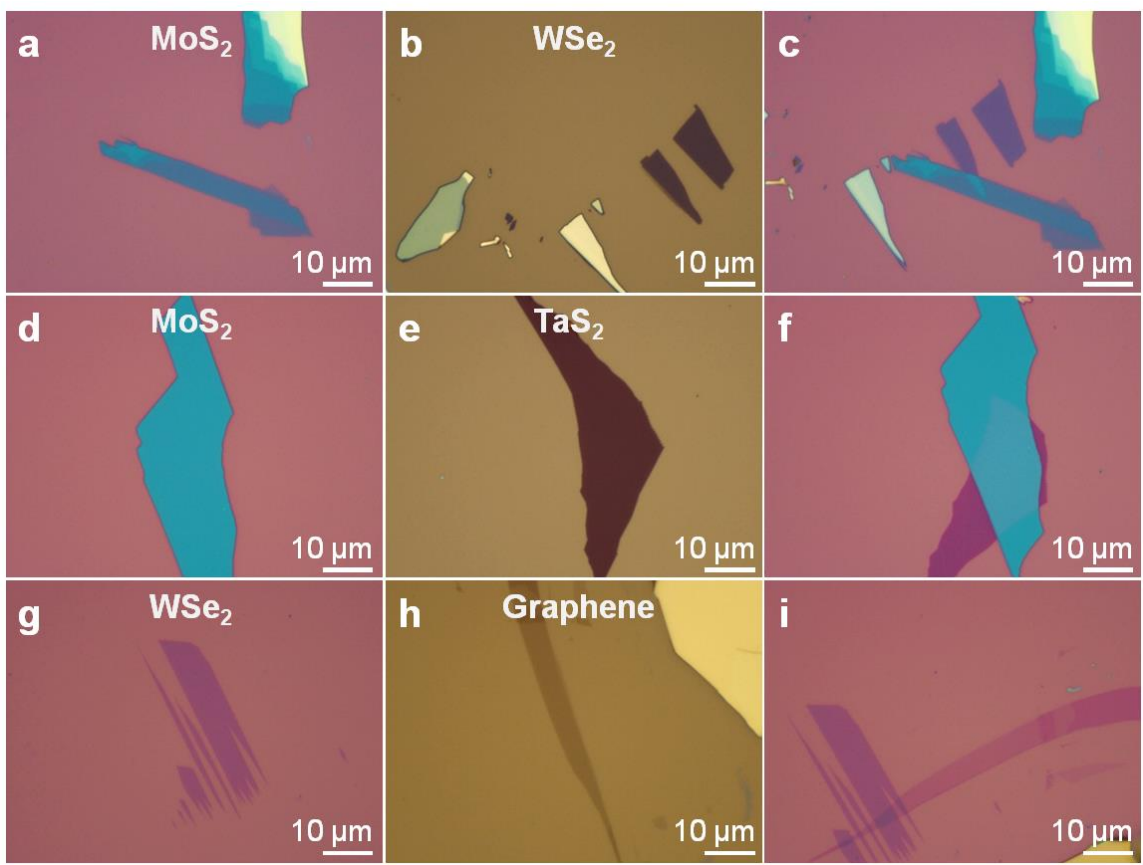

**Figure S11.** (a-c) OM images of (a) MoS$_2$, (b) WSe$_2$ nanosheets and (c) their heterostructure. (d-f) OM images of (d) MoS$_2$, (e) TaS$_2$ nanosheet and (f) their heterostructure. (g-i) OM images of (g) WSe$_2$, (h) graphene nanosheets and (i) their heterostructure.

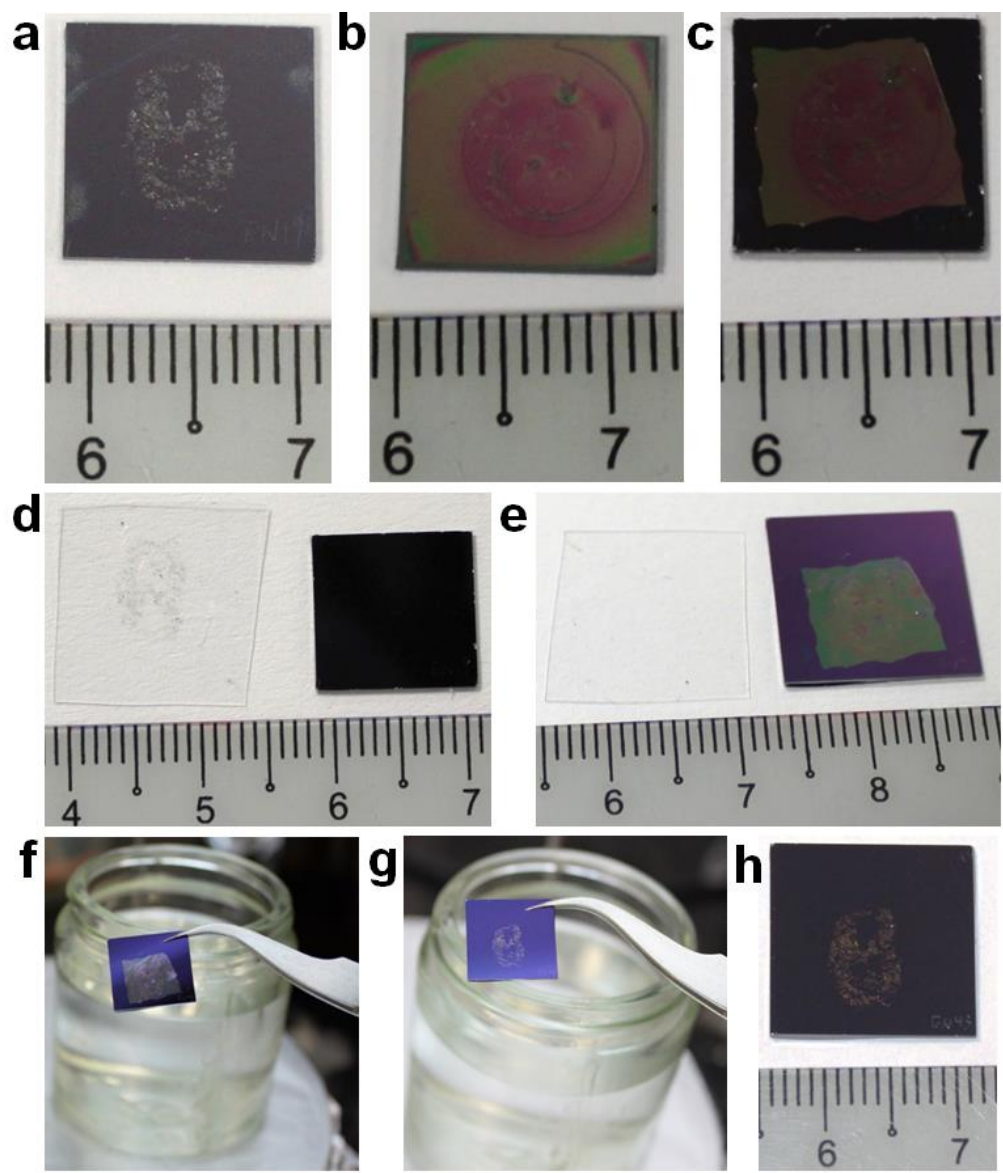

**Figure S12.** Photographs of the transfer and clean processes. MoS$_2$ flakes deposited on a 90 nm SiO$_2$/Si substrate with size of ca. 12 ×12 mm$^2$ before (a) and after (b) covered by the spin-coated PLLA film. (c) 1-mm-wide PLLA strips at the edges of the SiO$_2$/Si substrate have been removed to expose the hydrophilic SiO$_2$/Si. (d) The PLLA film and MoS$_2$ flakes are separated from the 90 nm SiO$_2$/Si substrate with a 1 to 2-mm thick PDMS film as mechanical support. (e) The PLLA film and MoS$_2$ flakes are transferred onto a 300 nm SiO$_2$/Si substrate with size of ca. 15 × 15 mm$^2$. The PLLA film and MoS$_2$ on 300 nm SiO$_2$/Si before (f) and after (g) being cleaned with DCM at 50 ºC. (h) The transferred MoS$_2$ flakes on the 300 nm SiO$_2$/Si target substrate.

**References:**

S1. Li, H.; Wu, J.; Qi, X.; He, Q.; Liusman, C.; Lu, G.; Zhou, X.; Zhang, H. Graphene Oxide Scrolls on Hydrophobic Substrates Fabricated by Molecular Combing and Their Application in Gas Sensing. *Small* **2013,** *9*, 382-386.
S2. Li, H.; Zhang, J.; Zhou, X. Z.; Lu, G.; Yin, Z. Y.; Li, G. P.; Wu, T.; Boey, F.; Venkatraman, S. S.; Zhang, H. Aminosilane Micropatterns on Hydroxyl-Terminated Substrates: Fabrication and Applications. *Langmuir* **2010,** *26*, 5603-5609.